# Free Radical Scavenging and Cytotoxic Activities of Substituted Pyrimidines

Qurat-ul-Ain,[1] Shafqat Hussain,[2] M. Iqbal Coudhary,[1,2,3] and Khalid Mohammed Khan[2,4]

[1]*Dr. Panjwani Center for Molecular Medicine and Drug Research, International Center for Chemical and Biological Sciences, University of Karachi, Karachi-75270, Pakistan*
[2]*H. E. J. Research Institute of Chemistry, International Center for Chemical and Biological Sciences, University of Karachi, Karachi-75270, Pakistan*
[3]*Department of Biochemistry, Faculty of Science, King Abdulaziz University, Jeddah-21589, Saudi Arabia*
[4]*Department of Clinical Pharmacy, Institute for Research and Medical Consultations (IRMC), Imam Abdulrahman Bin Faisal University, P.O. Box 1982, Dammam, 31441, Saudi Arabia*

iqbal.choudhary@iccs.edu[*]

**Summary:** A library of substituted pyrimidines was synthesized and evaluated for free radical scavenging, and *in vitro* cytotoxic activity in 3T3 cells. All compounds showed good free radical scavenging activity with IC$_{50}$ values in the range of 42.9 ± 0.31 to 438.3 ± 3.3 *μ*M as compared to the standard butylated hydroxytoluene having IC$_{50}$ value of 128.83 ± 2. 1 *μ*M. The structure activity-relationship was also established. Selected analogues **1**, **2**, **3**, **5**, **6**, **7**, **8**, **9**, **10**, **12**, **13**, **15**, **19**, **20**, **21**, **24**, **25**, **26** and **28** were tested for cytotoxicity in mouse fibroblast 3T3 cell line using MTT assay, and most of the analogues showed cytotoxicity. This study has identified a number of cytotoxic novel substituted pyrimidines having free radical scavenging activities that can be used as inhibitory compounds for those cancer cells whose growth is mediated by reactive oxygen species.

**Keywords:** Pyrimidine nucleotide, synthesis, free radical scavenging, SAR, cytotoxicity

**Introduction:**

Scavenging of free radicals by antioxidant compounds is an important biological function that may maintain in the body a low oxidative damage.[1] Antioxidant compounds of different synthetic, and natural sources can scavenge these free radicals with the formation of less reactive species, and thus diminish the radical induced oxidative damage that is possibly associated with many diseases, including cancers.[2-5] Numerous classes of synthetic compounds have been screened to reveal their free radical scavenging ability, including synthetically obtained deoxyribonucleic acids (DNA) and nucleotide analogues like pyrimidine derivatives.[6,7] These pyrimidines, present in numerous pharmaceutically important compounds, have been known to prevent cancer cell proliferation. Substituted pyrimidine primarily display their anticancer activity through intercalating with DNA nucleotide bases. However, they may prevent ROS induced DNA mutations in a way similar to other anticancer and antiviral molecules.[8-11] In recent years, anticancer drugs already being used in medical practice or being tested in clinical studies have been often based on pyrimidine skeleton, and new pyrimidine derivatives continue to show promising activities.[12-13] However, synthesis of antioxidant molecules can be a new approach to prevent proliferation of tumors whose growth is mediated by oxygen species.[14] Besides their anti- tumor action, pyrimidine derivatives have also been found to possess additional biological activities including antibacterial, anti-folate, antibiotic, anti-HIV, anti-fungal, anti-mycobacterial, anti-leismanial were also found to inhibit tumor necrotic factor alpha (TNF-α) production.[15] Herein, we report the free radical scavenging activities of a new library of pyrimidine derivatives to evaluate their potential against free radical sustained cancer cell proliferation. In the past, a number of pyrimidines were also found to inhibit enzymes such as tyrosine kinases, urease, *β*-glucuronidase, and cholinesterase.[16-18] Furthermore, many pyrimidine analogues were found to exhibit inhibitory or modulatory activities in a number of biological situations.[19, 20] Therefore, we screen these synthetic pyrimidine derivatives for their *in vitro* free radical scavenging activity as well as to establish their cytotoxicity in a 3T3 mouse fibroblast cell line.

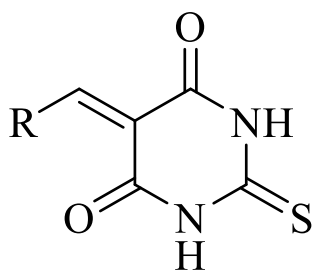

Scheme 1. Basic Skeleton of Pyrimidines.

**Experimental**
*Material and Methods*

All substituted pyrimidines were obtained from the in-house Molecular Bank facility of the Dr. Panjwani Center for Molecular Medicine and Drug Research, International Center for Chemical and Biological Sciences, University of Karachi, Pakistan. DPPH was purchased from Sigma Aldrich (Germany). Ethanol and dimethyl sulfoxide (DMSO) (reagent grade) were purchased from Sigma Aldrich (USA). Standard compounds, *i.e.,* butylated hydroxytoluene was purchased from Sigma Aldrich (Germany).

*DPPH Radical Scavenging Assay*

The Kumari Madhu method of DPPH (2,2-diphenyl-1-picryl-hydrazyl) assay[25] was used to measure the free radical scavenging activity with small variations. This assay is based on the reduction of DPPH radical (violet colour) by free radical scavenger with a change of colour to pale yellow. The intensity of colour conversion is directly related to the potency of free radical scavenging compounds, and to the extent of reduction in absorbance. In the visible region, absorbance reduction can be measure at 517 nm. Compounds solutions of (0.5 mM) in DMSO were prepared. Two-fold dilution method was used to dilute compounds solutions to different concentrations. 5 *µl* sample of each concentration was transferred to 96 wells plate in triplet, at 517 nm pre read was recorded. 95 *µ*l of 0.3 mM freshly prepared ethanolic solution of DPPH was added in each of the 96 wells. A final absorbance reading was taken at 517 nm. DMSO was used as negative control and butylated hydroxytoluene was used as the positive control. The radical scavenging activities were calculated by the following equation:

% Radical scavenging activity of DPPH
$$= [A_0 - A_1 / A_0] \times 100$$

Where:
$A_0$: The absorbance of all reagents without the tested compounds.
$A_1$: The absorbance in the presence of test compounds.

## MTT Assay

The pyrimidine derivatives were tested by the method previously described by Dimas *et al*. to establish their cytotoxicities in a normal cell line.[26] In 96-well plate, mouse fibroblast 3T3- cells ($2 \times 10^5$ cells/ mL) were grown over night in DMEM medium along with 10% FBS, pen/ strep (100 units/ mL), supplemented with 5% $CO_2$ at 37 °C. After 24 h, the old media was discarded, cells were treated with different concentrations of the tested compound, and further incubated for 24 h. After 24 h, cells were washed, and the plate was again incubated with MTT solution for 4 h after which dimethyl sulfoxide 100uL added for 15 min to dissolved formazan crystals at room temperature. Finally, a micro plate reader (SpectraMax Plus-384) was used to record the absorbance at 540 nm. The $IC_{50}$ was calculated and defined as the drug concentration ($\mu$M) causing cytotoxicity in 50%. cells.

## Results and Discussion
### Free Radical Scavenging Activity

The synthetic pyrimidine derivatives **1-28** were tested for their free radical scavenging, and cell cytotoxic potential. All compounds showed various degrees of radical scavenging activity inDPPH radical scavenging assay, and their $IC_{50}$ values ranged between $42.9 \pm 0.31$ to $438.3 \pm 3.3$ $\mu$M. Derivatives **1**, **3**, **11**, **13**, **18**, **26**, and **28** with $IC_{50}$ values of $55.6 \pm 2.1$, $122.4 \pm 1.9$, $107.65 \pm 1.3$, $108.4 \pm 2.8$, $113.4 \pm 1.3$, $42.9 \pm 0.31$, and $65.7 \pm 1.80$ $\mu$M, respectively, showed free radical inhibitory activity that is many folds better than the standard butylated hydroxytoluene with $IC_{50}$ value of $128.83 \pm 2.1$ $\mu$M, as depicted in Figs. A-D, and Table-1. Compounds **2**, **5**, **8**, **12**, **19**, and **27** showed good to moderate activities (Fig. A-C and Table-1). The remaining derivatives, including **6**, **7**, **9**,**10**, **15**, **21**, **22**, **23**, and **24** showed weak inhibitory activities (Fig. A-D and Table-1). Derivatives **4**, **14**, **16**, **17**, **20**, **25** were decleared as inactive derivatives of this series.

**Table-1:** Free radical scavenging activity of compounds (**1–28**).

| Cpds | IUPC Names | R | $IC_{50} \pm SEM^a$ ($\mu$M) |
|---|---|---|---|
| 1 | 5-(4-Hydroxy-3,5-dimethoxybenzylidene)-2-thioxodihydropyrimidine-4,6(1*H*,5*H*)-dione | 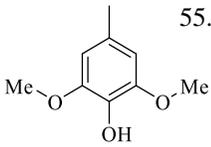 | $55.6 \pm 2.1$ |
| 2 | 5-(2-Bromo-4,5-dimethoxybenzylidene)-2-thioxodihydropyrimidine-4,6(1*H*,5*H*)-dione | 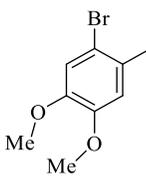 | $198.2 \pm 4.5$ |
| 3 | 5-((2-Hydroxynaphthalen-1-yl)methylene)-2-thioxodihydropyrimidine-4,6(1*H*,5*H*)-dione | 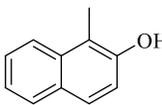 | $122.4 \pm 1.9$ |
| 4 | 5-(Thiophen-2-ylmethylene)-2-thioxodihydropyrimidine-4,6(1*H*,5*H*)-dione | 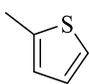 | NA |
| 5 | 2-Thioxo-5-(3,4,5-trimethoxybenzylidene)dihydropyrimidine-4,6(1*H*,5*H*)-dione | 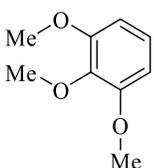 | $132.6 \pm 1.2$ |
| 6 | 5-(4-(Methylthio)benzylidene)-2-thioxodihydropyrimidine-4,6(1H,5H)-dione | 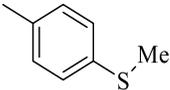 | 209 ± 4.4 |
| 7 | 5-((6-Bromo-4-chloro-2-oxo-2H-chromen-3-yl)methylene)-2-thioxodihydropyrimidine-4,6(1H,5H)-dione | 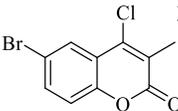 | 322.4 ± 1.9 |
| 8 | 5-(Pyridin-4-ylmethylene)-2-thioxodihydropyrimidine-4,6(1H,5H)-dione | 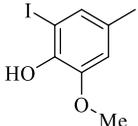 | 179.7 ± 6.2 |
| 9 | 5-((6-Methylpyridin-2-yl)methylene)-2-thioxodihydropyrimidine-4,6(1H,5H)-dione | 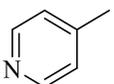 | 211.2 ± 4.6 |
| 10 | 5-(4-Bromo-2,5-dimethoxybenzylidene)-2-thioxodihydropyrimidine-4,6(1H,5H)-dione | 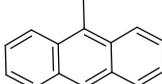 | 204.5 ± 3.5 |
| 11 | 5-(3-Hydroxy-4-methoxybenzylidene)-2-thioxodihydropyrimidine-4,6(1H,5H)-dione | 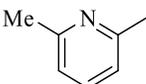 | 107.65 ± 1.3 |
| 12 | 5-(3,4-Dimethoxybenzylidene)-2-thioxodihydropyrimidine-4,6(1H,5H)-dione | 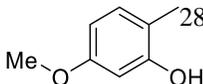 | 170.4 ± 2.5 |
| 13 | 5-(4-Hydroxy-3-iodo-5-methoxybenzylidene)-2-thioxodihydropyrimidine-4,6(1H,5H)-dione | 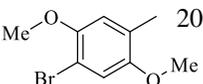 | 108.4 ± 2.8 |
| 14 | 5-(Anthracen-9-ylmethylene)-2-thioxodihydropyrimidine-4,6(1H,5H)-dione | 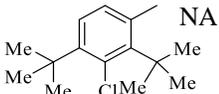 | NA |
| 15 | 5-(2-Hydroxy-4-methoxybenzylidene)-2-thioxodihydropyrimidine-4,6(1H,5H)-dione | 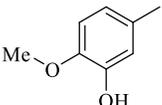 | 284.2 ± 5.9 |
| 16 | 5-(2,4-Di-tert-butyl-3-chlorobenzylidene)-2-thioxodihydropyrimidine-4,6(1H,5H)-dione | 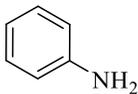 | NA |
| 17 | 5-(2-Aminobenzylidene)-2-thioxodihydropyrimidine-4,6(1H,5H)-dione | 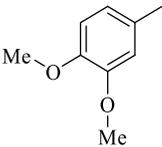 | NA |
| 18 | 5,5'-(1,4-Phenylenebis(methanylylidene))bis(2-thioxodihydropyrimidine-4,6(1H,5H)-dione) | 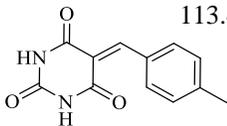 | 113.4 ± 1.1 |
| 19 | 5-(3,5-Dibromo-4-hydroxybenzylidene)-2-thioxodihydrop | 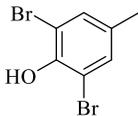 | 170.8 ± 1.4 |
| | yrimidine-4,6(1H,5H)-dione | | |
| 20 | 5-(4-(Dimethylamino)benzylidene)-2-thioxodihydropyrimidine-4,6(1H,5H)-dione | 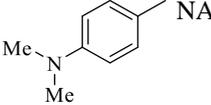 | NA |
| 21 | 5-(2-Methylbenzylidene)-2-thioxodihydropyrimidine-4,6(1H,5H)-dione | 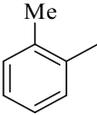 | 438.3 ± 3.3 |
| 22 | 5-(4-Ethoxybenzylidene)-2-thioxodihydropyrimidine-4,6(1H,5H)-dione | 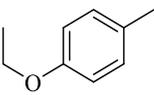 | 230.7 ± 2.6 |
| 23 | 5-(2,4-Dihydroxybenzylidene)-2-thioxodihydropyrimidine-4,6(1H,5H)-dione | 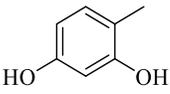 | 231.9 ± 6.9 |
| 24 | 5-(2-Hydroxy-3-methoxybenzylidene)-2-thioxodihydropyrimidine-4,6(1H,5H)-dione | 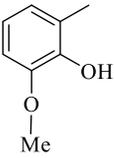 | 200.6 ± 1.8 |
| 25 | 5-((5-Methylfuran-2-yl)methylene)-2-thioxodihydropyrimidine-4,6(1H,5H)-dione | 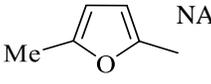 | NA |
| 26 | 5-(3,4-Dihydroxybenzylidene)-2-thioxodihydropyrimidine-4,6(1H,5H)-dione | 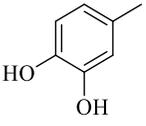 | 42.9 ± 3.6 |
| 27 | 5-(2-Hydroxy-5-methoxybenzylidene)-2-thioxodihydropyrimidine-4,6(1H,5H)-dione | 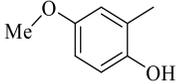 | 177.1 ± 3.6 |
| 28 | 2-Thioxo-5-(2,3,4-trihydroxybenzylidene)dihydropyrimidine-4,6(1H,5H)-dione | 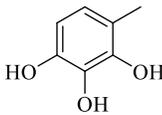 | 65.7 ± 1.8 |
| BHT[b] | 2,6-Di-tert-butyl-4-methylphenol | 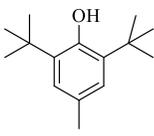 | 128.8 ± 2.1 |

[a]SEM is the standard error of the mean, BHT[b]: Butylated hydroxytoulene

## Structure- Activity Relationship

A structure-activity relationship established for all compounds that confirmed substitution of various functionalities at the aromatic ring confers free radical scavenging activity to each particular pyrimidine analogue. Analogue **26**, a 3,4-dihydroxybenzylidene was found to be the most active pyrimidine among the series, with an IC$_{50}$ value of 42.9 ± 0.31 μM, corresponding to 84.07% radical scavenging activity that is as good as 85.87% radical scavenging activity of the standard drug (Tables-1, and 3). The high activity shown by analogue **26** is due to the positional change of dihydroxyl groups present an aromatic moiety (Table-1). Literature reports have also shown that the phenolic hydroxyl group is responsible for the antioxidant function.[21,24] Compound **1** is the second most potent derivative among the series, containing 4-hydroxy-3,5-dimethoxy groups with IC$_{50}$ of value 55.6 ± 2.1 μM, with corresponding 91.58% radical scavenging activity (Tables-1, and 3). With 74.69% radical scavenging activity derivative **28**, with three hydroxyl groups at 2,3, and 4-positions, was found to be the third most effective derivative of the series (Tables-1, and 3). The lesser activity shown by analogue **28** as compared to compound **26** might be due to the extra hydroxyl group which creates some steric hindrance (Table-1 and 3). In this study, we observed that all other hydroxyl group containing derivatives, such as **3**, **11**, **13**, **15**, **19**, **23** and **27,** also showed antioxidant activity. The lesser activity shown by analogue **28** as compared to compound **26** might be due to the extra hydroxyl group which creates some steric hindrance (Table-1 and 3). In this study, we observed that all other hydroxyl group containing derivatives, such as **3**, **11**, **13**, **15**, **19**, **23** and **27,** also showed antioxidant activity. The difference in their activity seems to be either due to the number, position, and presence of other substituents along with the hydroxyl group (Table-1). Compound **8**, and **12** have almost identical free radical scavenging activity with 74.3%, and 90.76 % (Table-3). The moderate activity of compound **8** may be due to the lone pair of electrons on the pyridine nitrogen while in derivative **12**, due to the presence of two methoxy groups (Table-1). Compound **7**, and **20** with (6-bromo-4-chloro-2-oxo-2*H*-chromen-3-yl) and (6-bromo-4-chloro-2-oxo-2*H*-chromen-3-yl) substitutions were found to be the least active of the series (Tables-1, and 3). The anthranyl analogue **14**, di *tert*-butyl compound **16**, derivative **17** having aminobenzylidene, derivative **20** with dimethylamino group, methylfuryl molecule **25**, and thiophenyl derivative **4** did not show any antioxidant activity. 4-Bromo-2,5-dimethoxy compound **10** and 2-bromo-4,5-dimethoxy analogue **2** have the same substituents but their positions are different, providing little difference in their activities (Table-1). By changing the substituent from *p*-thiomethyl, as in analogue **6**, to an amino groups such as *N*, *N*-dimethyl amino derivative **19** and methyl-2-pyridinyl molecule **9**, it was observed the amino analogues showed greater radical scavenging activity than the one with *p*-thiomethyl and *N*, *N*-dimethyl amino functionalities. This might be due to the better ability of the former to provide free electrons (Tables-1, and 3).

*Cell Cytotoxic Activity*

Cytotoxicity of compounds **1**, **2**, **3**, **5**, **6**, **7**, **8**, **9**, **10**, **12**, **13**, **15**, **19**, **20**, **21**, **24**, **25**, **26** and **28** was carried out by using mouse fibroblast 3T3 cell line. Derivatives **1**, **3**, **5**, **6**, **8**, **9**, **12**, **13**, **15**, **19**, **21**, **24**, **25**, **26** and **28** exhibited non-cytotoxicity in mouse fibroblast 3T3 cell line (Table-3). Derivatives **7** and **20** were found to have weak cytotoxic effect with IC$_{50}$ values of 27.038 ± 0.26, and 22.4 ± 0.76, $\mu$M, respectively. However, compound **2** was found to be moderately cytotoxic with IC$_{50}$ value of 19.482 ± 0.406 $\mu$M, and only compound **10** was found to be cytotoxic with IC$_{50}$ value of 7.038 ± 0.26 $\mu$M.

**Table-2:** Cytotoxicity studies of selected pyrimidine derivatives.

| Compounds | Cell Cytotoxicity IC$_{50}$ ($\mu$M) ± SEM | Compounds | Cell Cytotoxicity IC$_{50}$ ($\mu$M) ± SEM |
|---|---|---|---|
| 1 | >30 | 12 | >30 |
| 2 | 19.4 ± 0.4 | 13 | >30 |
| 3 | >30 | 15 | >30 |
| 5 | >30 | 19 | >30 |
| 6 | >30 | 20 | 22.4 ± 0.7 |
| 7 | 27.0 ± 0.2 | 21 | >30 |
| 8 | >30 | 24 | >30 |
| 9 | >30 | 25 | >30 |
| 10 | 7.0 ± 0.2 | 26 | >30 |
| Cycloheximide | 0.2± 0.1 | 28 | >30 |

SEM: standard error mean, cycloheximide standard drug

**Table-3:** % RSA of selected derivative.

| Cpds | % Radical Scavenging Activity | Cpds | % Radical Scavenging Activity |
|---|---|---|---|
| **1** | 91.58 | **15** | 70.64 |
| **2** | 89.25 | **16** | 2.61 |
| **3** | 89.85 | **17** | 1.52 |
| **4** | 13.45 | **18** | 75.84 |
| **5** | 90.43 | **19** | 81.95 |
| **6** | 78.00 | **20** | 19.29 |
| **7** | 62.01 | **21** | 53.44 |
| **8** | 74.35 | **22** | 70.78 |
| **9** | 73.81 | **23** | 81.96 |
| **10** | 75.00 | **24** | 81.72 |
| **11** | 79.85 | **25** | 15.54 |
| **12** | 90.76 | **26** | 84.07 |
| **13** | 89.95 | **27** | 81.36 |
| **14** | 34.92 | **28** | 74.69 |

% RSA: % Radical Scavenging Activity, Butylated hydroxytoulene (BHT) % RSA: 85.87

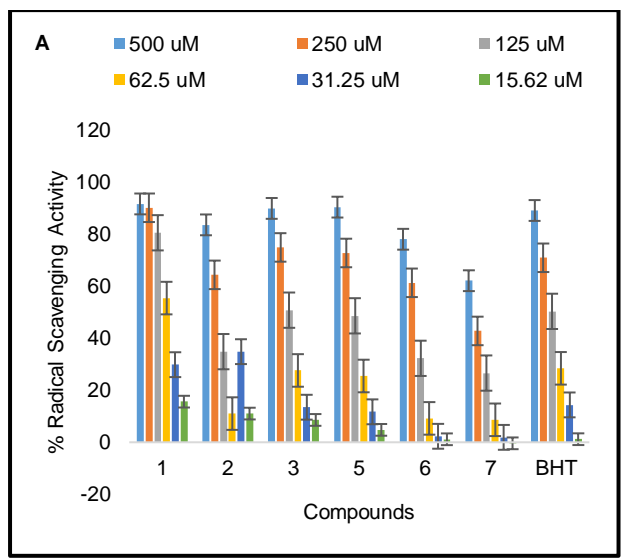
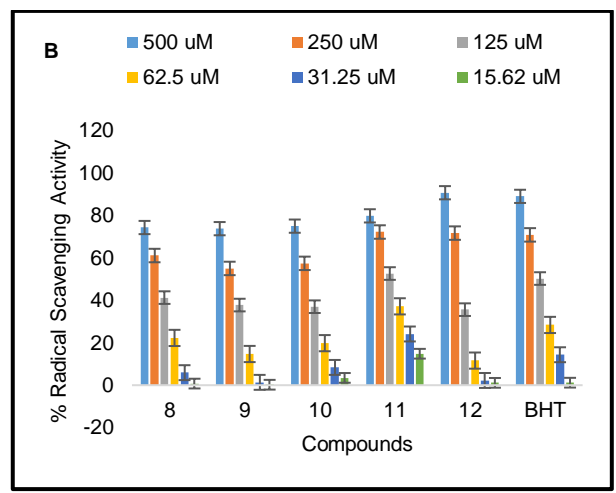
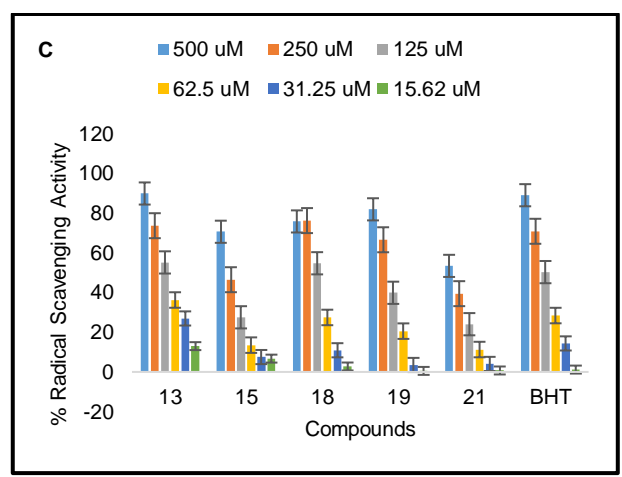
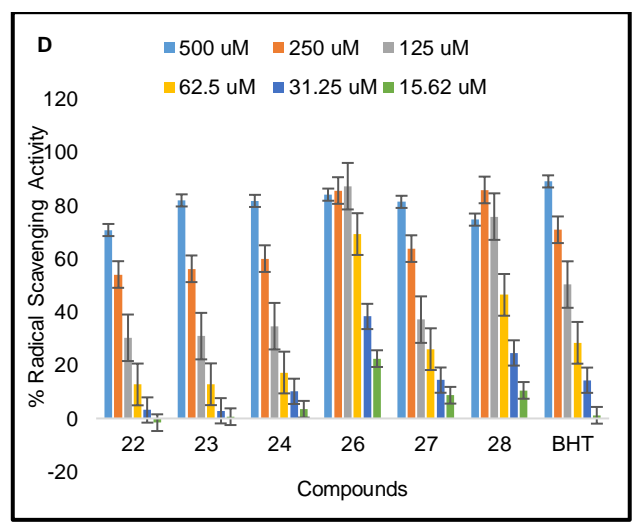

Figs. A-D: Free radical scavenging activities of pyrimidine derivatives (**1–28**) on deferent concentrations.

## Conclusion

The present study identifies new series of pyrimidines as potential radical scavengers. All analogues were found to display diverse free radical scavenging potential when compared with the standard butylated hydroxytoluene. Compounds **1**, **3**, **11**, **13**, **17**, **25,** and **27**, with IC$_{50}$ values of 55.6 ± 2.1, 122.4 ± 1.9, 107.65 ±1.3, 108.4 ± 2.8, 113.4 ±1.3, 42. 9± 0.31, and 65.7 ± 1.80 $\mu$M, respectively, showed good free radical scavenging potential better than the standard butylated hydroxytoluene having IC$_{50}$ value of 128.83 ± 2.1$\mu$M. Cytotoxic evaluation of selected derivatives further support our study. Compound **1**, **13**, and **25** were identified as non-cytotoxic against 3T3 cells; therefore, these can serve as lead compounds for further development as potential drug candidates to scavenge reactive oxygen species.

## Acknowledgments

The authors are thankful to the Pakistan Academy of Sciences for providing financial support to Project No. (5-9/PAS/440).

## References:

1. Neha, K., Haider, M. R., Pathak, A., and Yar, M. S., Medicinal prospects of antioxidants: A review. *European Journal of Medicinal Chemistry*, **178**, 687-704 (**2019**).
2. Suleman, M., Khan, A., Baqi, A., Kakar, M. S., and Ayub, M., Antioxidants, its role in preventing free radicals and infectious diseases in human body. *Pure and Applied Biology*, **8**, 380-388 (**2019**).
3. Young, A., and Lowe, G., Carotenoids-antioxidant properties. *Antioxiants*, **7**, 28, (**2018**).
4. Qurat-ul-Ain., Choudhary, M.I., and Kochanek, K.S., Modulation of melanoma cell proliferation and spreading by novel small molecular weight antioxidants. *Free Radical Biology and Medicine*, **108**, S28 (**2017**).
5. Pisoschi, A. M., and Pop, A., The role of antioxidants in the chemistry of oxidative stress: A review. *European Journal of Medicinal Chemistry*, **97**, 55-74 (**2015**).
6. Ain, Qurat-ul., Study of the Effect of Antioxidant on Oxidative Stress in Molecular and Cellular Models. (Doctoral dissertation, University of Karachi, Karachi), 1-227(**2019**).
7. Sankarganesh, M., Revathi, N., Raja, J. D., Sakthikumar, K., Kumar, G. G. V., Rajesh, J., and Mitu, L., Computational, antimicrobial, DNA binding and anticancer activities of pyrimidine incorporated ligand and its copper (II) and zinc (II) complexes (II) complexes. *Journal of Serbian Chemical Society*, **84**, 277-291(**2019**).
8. Haleel, A. K., Rafi, U. M., Mahendiran, D., Mitu, L., Veena, V., and Rahiman, A. K., DNA profiling and in vitro cytotoxicity studies of tetrazolo [1,5-a] pyrimidine-based copper (II) complexes. *Biometals*, 1-16 (**2019**).
9. Shringare, S. N., Chavan, H. V., Bhale, P. S., Dongare, S. B., Mule, Y. B., Kolekar, N. D., and Bandgar, B. P., Synthesis and pharmacological evaluation of pyrazoline and pyrimidine analogs of combretastatin-A4 as anticancer, anti-inflammatory and antioxidant agents. *Croatica Chemical Acta*, **91**, 1-10 (**2018**).
10. Wang, Z., Kang, D., Chen, M., Wu, G., Feng, D., Zhao, T., and Daelemans, D., Design, synthesis, and antiviral evaluation of novel hydrazone-substituted thiophene [3, 2-d] pyrimidine derivatives as potent human immunodeficiency virus-1 inhibitors.

*Chemical Biology and Drug Design*, **92**, 2009-2021(**2018**).

11. Kumar, S., Deep, A., and Narasimhan, B., A Review on Synthesis, Anticancer and Antiviral Potentials of Pyrimidine Derivatives. *Current Bioactive Compounds*, **15**, 289-303(**2019**).
12. Selvam, T. P., James, C. R., Dniandev, P. V., and Valzita, S. K., "A mini review of pyrimidine and fused pyrimidine marketed drugs. *Journal of Research in Pharmacy Practice*, **2**, 01-09 (**2015**).
13. Tokunaga, S., Takashima, T., Kashiwagi, S., Noda, S., Kawajiri, H., Tokumoto, M., and Mizuyama, Y. Neoadjuvant Chemotherapy with Nab-paclitaxel Plus Trastuzumab Followed by 5-Fluorouracil/Epirubicin/Cyclophosphamide for HER2-positive Operable Breast Cancer: A Multicenter Phase II Trial. *Anticancer Research*, **39**, 2053-2059 (**2019**).
14. Ghorab, M. M., and Alsaid, M. S. Anticancer activity of some novel thieno [2, 3-d] pyrimidine derivatives. *Biomedical Research*, **27**, 110-115 (**2016**).
15. Wilhelm, M., Mueller, L., Miller, M. C., Link, K., Holdenrieder, S., Bertsch, T., and Birkmann, J. Prospective, multicenter study of 5-fluorouracil therapeutic drug monitoring in metastatic colorectal cancer treated in routine clinical practice. *Clinical Colorectal Cancer*, **15**, 381-388 (**2016**).
16. Smaill, J. B., Gonzales, A. J., Spicer, J. A., Lee, H., Reed, J. E., Sexton, K., and Denny, W. A., Tyrosine kinase inhibitors. 20. Optimization of substituted quinazoline and pyrido [3, 4-d] pyrimidine derivatives as orally active, irreversible inhibitors of the epidermal growth factor receptor family. *Journal Medicinal Chemistry*, **59** (17), 8103-8124 (**2016**).
17. Khan, K. M., Rahim, F., Khan, A., Shabeer, M., Hussain, S., Rehman, W., and Choudhary, M. I., Synthesis and structure–activity relationship of thiobarbituric acid derivatives as potent inhibitors of urease. *Bioorganic and Medicinal Chemistry*, **22**, 4119-4123 (**2014**).
18. Barakat, A., Islam, M. S., Al-Majid, A. M., Ghabbour, H. A., Yousuf, S., Ashraf, M., and Ul-Haq, Z., Synthesis of pyrimidine-2,4,6-trione derivatives: Anti-oxidant, anti-cancer, α-glucosidase, *β*-glucuronidase inhibition and their molecular docking studies. *Bioorganic Chemistry*, **68**, 72-79 (**2016**).
19. Gobbi, L., Grether, U., Guba, W., Kretz, J., Martin, R. E., Westphal, M. V., and I. Jzerman, A. P. *U.S. Patent Application No. 16/228,543*, 1-13 (**2019**).
20. Reddy, E. K., Remya, C., Sajith, A. M., Dileep, K. V., Sadasivan, C., and Anwar, S., Functionalized dihydroazo pyrimidine derivatives from Morita-Baylis-Hillman acetates: synthesis and studies against acetylcholinesterase as its inhibitors. *RSC Advances*, **6**, 77431-77439 (**2016**).
21. Sahu, M., and Siddiqui, N., A review on biological importance of pyrimidines in the new era. *Journal of Pharmacy and Pharmaceutical Sciences*, **8**, 8-21(**2016**).
22. Ke, S., Shi, L., Zhang, Z., and Yang, Z., Steroidal [17, 16-d] pyrimidines derived from dehydroepiandrosterone: A convenient synthesis, antiproliferation activity, structure-activity relationships, and role of heterocyclic moiety. *Scientific Reports*, **7**, 44439 (**2017**).

23. Khan, K.M., Rahim, F., Shabeer, M., Khan, A., Hussain, S., Rehman, W., Taha, M. Khan, M., Perveen, S. and Choudhary, M. I., Synthesis and structure-activity relationship of thiobarbituric acid derivatives as potent inhibitors of urease. *Bioorganic and Medicinal Chemistry*, **22**, 4119-4123 (**2014**).
24. Priyadarsini, K. I., Maity, D. K., Naik, G. H., Kumar, M. S., Unnikrishnan, M. K., Satav, J., Mohan G.H., Role of phenolic OH and methylene hydrogen on the free radical reactions and antioxidant activity of curcumin. *Free Radical Biology Medicine*, **35**, 475-484 (**2003**).
25. Madhu, K., Phytochemical screening and antioxidant activity of in vitro grown plants *Clitoria ternatea* L., using DPPH assay. *Asian Journal of Pharmaceutical Clinical Research*, **6**, 38-42 (**2013**).
26. Vemana, H. P., Barasa, L., Surubhotla, N., Kong, J., Ha, S. S., Palaguachi, C., and Dukhande, V. V. Benzimidazole scaffolds as potential anticancer agents: Synthesis and Biological evaluation. *The FASEB Journal*, **33**, 646-18 (**2019**).